\documentclass[aps,twocolumn,prl,showpacs,superscriptaddress,groupedaddress]{revtex4}
\usepackage{epsfig}\usepackage{amssymb}
\usepackage{amsmath}
\usepackage{amsfonts}
\usepackage{graphicx}
\usepackage{mathrsfs}
\usepackage{multirow}
\usepackage{wrapfig}
\usepackage{hyperref}
\usepackage{mathtools}
\usepackage{blindtext}
\usepackage{dcolumn}   

\newcommand{\bC}{\mathbf{C}}

\newcommand{\cJ}{\mathcal{J}}

\newcommand{\be}{\begin{equation}}
\newcommand{\ee}{\end{equation}}
\newcommand{\bea}{\begin{eqnarray}}
\newcommand{\eea}{\end{eqnarray}}

\newcommand{\ed}{\end{document}}

\newcommand{\bi}{\begin{itemize}}
\newcommand{\ei}{\end{itemize}}

\newcommand{\bce}{\begin{center}}
\newcommand{\ece}{\end{center}}

\DeclareUnicodeCharacter{2212}{-}

\begin{document}

\title{Exploring Metamaterial Lasers through Non-Hermitian Scattering Formalism}

\author{Özge Beyza Vardar}\email{vardar.ozge11@gmail.com}\affiliation{Institute of Graduate Studies in Science, Istanbul University, Istanbul 34134, Türkiye}
\author{Uğur Tamer}\email{ugurtamerphysics@gmail.com}\affiliation{Department of Physics, Istanbul University, 34134, Vezneciler,
Istanbul, Türkiye}
\author{Mohammad Mehdi Sadeghi}\email{sadeghi@istanbul.edu.tr}\affiliation{Department of Physics, Istanbul University, 34134, Vezneciler,
Istanbul, Türkiye}
\author{Mustafa Sarısaman}\email{mustafa.sarisaman@istanbul.edu.tr}\affiliation{Department of Physics, Istanbul University, 34134, Vezneciler,
Istanbul, Türkiye}
\affiliation{National Intelligence Academy, Institute of Engineering and Science, Ankara, Türkiye}

\begin{abstract}
 This study explores the exciting properties of metamaterials and their innovative applications in non-Hermitian physics, with particular emphasis on the scattering formalism, a key topic of recent research. We have analyzed how light behaves in a negative index metamaterial (NIM), allowing us to develop a transfer matrix and identify the essential conditions for the occurrence of spectral singularities. These findings are crucial for fine-tuning system parameters that will drive the development of metamaterial slab lasers and coherent perfect absorber (CPA) systems. Overall, our research demonstrates the enormous potential of metamaterials and their significant role in driving innovation in various technology areas.
\end{abstract}


\maketitle

\section{Introduction}

Metamaterials are engineered materials that go beyond the capabilities of natural substances in manipulating electromagnetic waves. Pioneered by scientists like Veselago \cite{veselago} and Pendry \cite{pendry2000}, these materials are designed using sub-wavelength structures that enable properties not found in nature, such as negative permittivity and permeability. This leads to extraordinary phenomena, including negative refraction \cite{eleftheriades2005, agranovich2004, smith2004, hoffman2007, fang2009, Litchinitser2008} and perfect lenses \cite{rosenblatt2015, zharov2005, zhang2011}.

Veselago’s groundbreaking work in the 1960s showed that materials with both negative permittivity and permeability could result in a negative refractive index, making light bend in the opposite direction from how it behaves in conventional materials \cite{veselago}. This unusual behavior sparked the exploration of materials with these counter-intuitive properties. Pendry, in 2000, took the concept further by introducing the idea of metamaterials—artificially structured materials designed to achieve properties that are otherwise unattainable in nature \cite{pendry2000}. Pendry showed that by manipulating the geometry of materials at sub-wavelength scales, we could design materials with tailored electromagnetic properties, including negative values for permittivity and permeability.

Since then, the field of metamaterials has expanded rapidly, unlocking new possibilities for controlling light and other forms of electromagnetic radiation. These materials are already being explored for use in super-resolution imaging, which could revolutionize microscopy and medical diagnostics by overcoming the diffraction limit \cite{liu2007}. Metamaterials could also be used to create cloaking devices that bend light around objects, rendering them invisible, or to develop radar and electromagnetic shielding technologies with improved performance \cite{schurig2006, holloway2005}. In addition, they hold promise for enhancing the design of antennas and improving communication systems. 

Metamaterials can also be used in acoustic applications, where they can manipulate sound waves in unprecedented ways. For example, they could lead to acoustic cloaks that make objects soundproof or even focus sound waves like lenses, offering new solutions for noise cancellation and advanced sound control \cite{cummer2004}.

The potential of metamaterials is vast, and the field continues to evolve with new directions. One exciting development is the integration of artificial intelligence (AI) to optimize the design of metamaterials \cite{zhu2017}. AI could help accelerate the discovery of novel structures and materials with enhanced properties that would be difficult to design using traditional methods. Another promising area is combining metamaterials with other advanced materials, such as graphene or topological insulators, to create multifunctional materials that can enable cutting-edge technologies like spintronics, quantum computing, and energy harvesting \cite{cai2010, spint1, spint2, spint3}.

Additionally, researchers are looking to nature for inspiration, studying structures like butterfly wings and spider silk to design bio-inspired metamaterials that are lightweight, biocompatible, and suitable for medical applications or soft robotics \cite{liu2010}.

As research progresses, the potential applications of metamaterials will continue to expand. From manipulating light and sound to enhancing communication systems and creating advanced medical technologies, the future of metamaterials holds transformative possibilities that could reshape how we interact with the physical world.

One of the possible uses of metamaterials is in the field of lasers\cite{tavallaee2010, ziolkowski2006, boardman2011, fu2017}. Laser physics is a phenomenon that can be understood best by the existence of spectral singularities. Spectral singularities are a phenomenon that generates waves in a system that completely exit the system, making them closely related to the principles of laser physics\cite{prl-2009, p123}. We point out that spectral singularity points may be considered in a physical system with a continuous spectrum\cite{naimark, naimark-1, p123}. With this in mind, the interaction of metamaterials with electromagnetic waves is actually an electromagnetic scattering theory with non-Hermitian character\cite{CPA, Oktay2020}.

Studying metamaterials with the tools of non-Hermitian physics is a very intriguing and notable approach \cite{mostafazadehmet1}. It is customary to know that Hermitian systems have real eigenvalues and orthonormal eigenstates. On the other hand, non-Hermitian physical systems may involve exceptional points, unlike Hermitian systems\cite{bender, ijgmmp-2010, longhi4, longhi3, nonhermit1, nonhermit2, nonhermit3, nonhermit4, nonhermit5, nonhermit6, nonhermit7, nonhermit8, nonhermit9, nonhermit10, nonhermit11, nonhermit12, nonhermit13}. The system may have real eigenvalues at these exceptional points, while the eigenstates coalesce. Spectral singularities of any optical system in this picture, in case of scattering, correspond to appear states of divergent reflection and transmission amplitudes for the real $k$ values of the physical system\cite{p123, prl-2009, CPA, lastpaper, pra-2011a, pra-2012a, mostafazadehmet2}, where $k$ is the wave number. This causes the zero-width resonance and the laser threshold state to occur, as it generally produces purely outgoing waves\cite{prl-2009}. This is a natural consequence of non-Hermitian physics, unlike traditional lasers. In recent years, very essential studies have been carried out in understanding many unknown aspects of new phenomena and realities with non-Hermitian physics, and voluminous intriguing studies are currently being carried out\cite{nonhermit1, nonhermit2, nonhermit3, nonhermit4, nonhermit5, nonhermit6, nonhermit7, nonhermit8, nonhermit9, nonhermit10, nonhermit11, nonhermit12, nonhermit13}. The non-Hermitian physics thus plays a crucial role in understanding the exotic properties of the metamaterials\cite{Oktay2020, hamed2020, sarisaman20192}. This idea constitutes the main motivation of our work. In this study, we will focus on negative index metamaterials (NIMs) and perform our analysis on these types of metamaterials. However, our results are also valid for positive index metamaterials.

\section{A. TE/TM Mode Configurations of a Negative Index Metamaterial Slab }
\label{S2}

\begin{figure*}
\centering
\includegraphics[width=8cm]{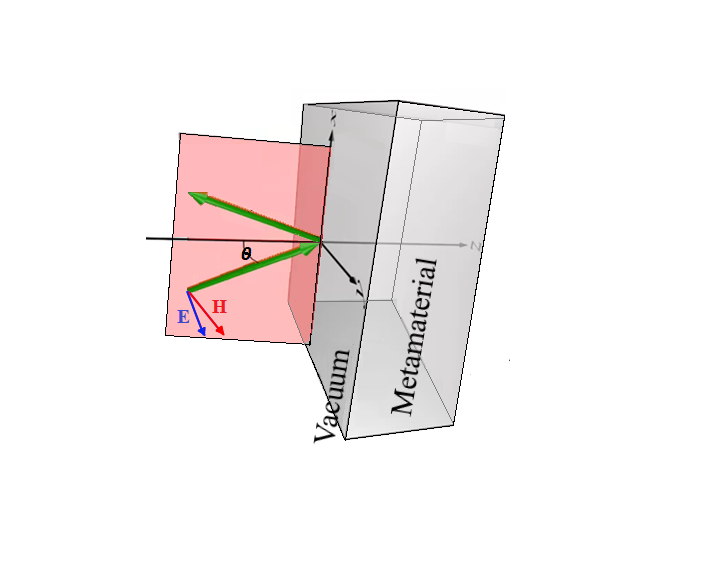}
\includegraphics[width=8cm]{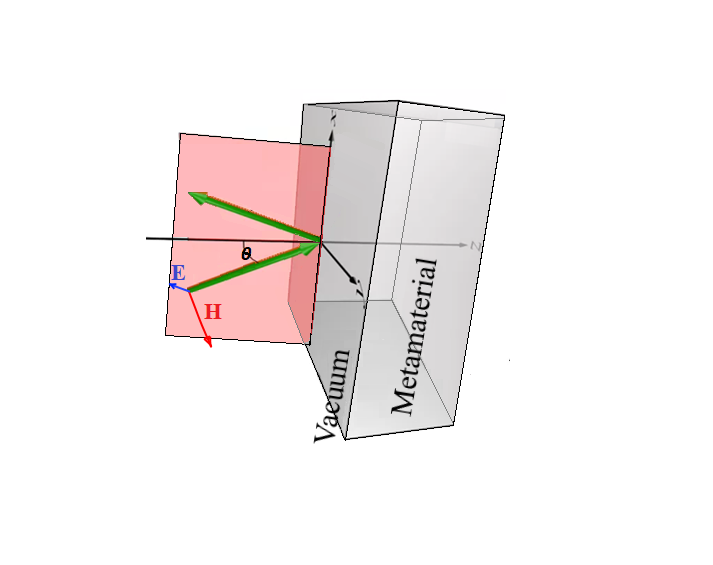}
    \caption{(Color Online) Figure displays the TE (left panel) and TM (right panel) mode configurations for the interaction of electromagnetic wave to the (NIM) metamaterial environment. Wave is emitted by an angle $\theta$ from the normal to the surface with the metamaterial slab system. Plane of incidence is normal to the surface, and is shown by reddish color.}
    \label{fig1}
\end{figure*}

Consider an optically active negative index metamaterial slab system with thickness $L$, aligned along the $z$-axis as shown in Fig.\ref{fig1}. We let the slab possess a complex refractive index, which is uniform between the end-faces in region $z\in [0, L]$, i.e. (NIM) metamaterial has a compact support. Recall that complex refractive index is evaluated in terms of permittivity and permeability of the medium as $\textbf{n}:=1/\sqrt{\epsilon\mu}$. The interaction of this distinctively equipped material with electromagnetic waves is very important for understanding its scattering properties and basic characteristics. Thus one may realize its applications in technologies of quantum devices. Notice that artificially fabricated metamaterial allows a broad range of refractive indices, not only in positive but also in negative domains. The governing equations corresponding to the wave propagation in a metamaterial medium provided by the Drude Model\cite{Li2007, Li2008, Huang2012, Rihani2022} are

\begin{align}
&\vec{\nabla}\times\vec{H}= \vec{J} + \varepsilon_{0}\frac{\partial\vec{E}}{\partial t},\notag\\
&\vec{\nabla}\times\vec{E}= - \vec{K} -\mu_{0}\frac{\partial\vec{H}}{\partial t},\notag\\
&\frac{\partial \vec{J}}{\partial t} + \Gamma_e \vec{J} = \varepsilon_{0} \omega_{pe}^2 \vec{E},\notag\\
&\frac{\partial \vec{K}}{\partial t} + \Gamma_m \vec{K} = \mu_{0} \omega_{pm}^2 \vec{H},\label{maxwell}
\end{align}
where $\varepsilon_{0}$ and $\mu_{0}$ are vacuum permittivity and permeability, respectively; $\omega_{pe}$ and $\omega_{pm}$ are electric and magnetic plasma frequencies, respectively; $\Gamma_e$ and $\Gamma_m$ are the electric and magnetic damping frequencies, respectively; $\vec{J}(\vec{r}, t)$ and $\vec{K}(\vec{r}, t)$ are the induced electric and magnetic currents, respectively; and finally $\vec{E}(\vec{r}, t)$ and $\vec{H}(\vec{r}, t)$ are the electric and magnetic fields, respectively. We may express the constitutive relations regarding $\vec{D}$ and $\vec{B}$ fields as follows
\begin{align}
\vec{D} (\vec{r}, t) &:= \varepsilon \vec{E} (\vec{r}, t)= \varepsilon_0 \vec{E} (\vec{r}, t) + \vec{P} (\vec{r}, t),\label{Dfield} \\
\vec{B} (\vec{r}, t) &:= \mu \vec{H} (\vec{r}, t) = \mu_0 \vec{H} (\vec{r}, t) + \vec{M} (\vec{r}, t),\label{Bfield}
\end{align}
where $\vec{P}$ and $\vec{M}$ vectors represent the electric and magnetic polarization vectors. These vectors are associated with induced currents $\vec{J}$ and $\vec{K}$ as follows
\be
\vec{J} := \frac{\partial \vec{P}}{\partial t}\qquad \vec{K} := \frac{\partial \vec{M}}{\partial t}, 
\ee
Time harmonic waves yield a field representation in time domain by $\vec{\Psi}(\vec{r}, t)=e^{-i\omega t} \vec{\Psi}(\vec{r})$, where $\vec{\Psi}$ implies all time-dependent fields in our setup. Therefore, one obtains the following governing equations in the frequency domain
\begin{align}
\vec{\nabla}\times\vec{H}&= \vec{J} -i\omega\varepsilon_{0}\vec{E},\label{curlH2}\\
\vec{\nabla}\times\vec{E}&= - \vec{K} + i\omega\mu_{0}\vec{H},\label{curlE2}\\
\vec{J} &= \frac{\varepsilon_{0} \omega_{pe}^2}{\Gamma_e -i\omega}\vec{E},\label{J2}\\
\vec{K} &= \frac{\mu_{0} \omega_{pm}^2}{\Gamma_m -i\omega}\vec{H},\label{K2}
\end{align}
In these equations one directly realizes the following symmetry relations between Eqs.~(\ref{curlH2}) and (\ref{curlE2}), and Eqs.~(\ref{J2}) and (\ref{K2})
\begin{align}
\vec{E}\rightarrow &\vec{H}, \qquad \vec{H}\rightarrow -\vec{E},\qquad
\vec{J}\rightarrow\vec{K}, \qquad \vec{K}\rightarrow -\vec{J},\notag\\
&\varepsilon_{0}\longleftrightarrow \mu_0,\quad \omega_{pe} \longleftrightarrow \omega_{pm},\quad \Gamma_{e} \longleftrightarrow \Gamma_{m}. \notag
\end{align}
In fact, $\vec{E} \& \vec{H}$ and  $\vec{J} \& \vec{K}$ field couples are transformed into each other by means of two-dimensional anti-symmetric Levi-Civita symbol $\epsilon^{\alpha\beta}$ respecting the order such that $\vec{E}^{\alpha} \xleftrightarrow[\epsilon^{\beta\alpha}]{\epsilon^{\alpha\beta}}\vec{H}^{\beta}$ and $\vec{J}^{\alpha} \xleftrightarrow[\epsilon^{\beta\alpha}]{\epsilon^{\alpha\beta}}\vec{K}^{\beta}$ with $\epsilon^{\beta\alpha} = -\epsilon^{\alpha\beta}$. On the other hand, we notice that electric and magnetic polarization vectors are determined to be
\be
\vec{P} = -\frac{\varepsilon_{0} \omega_{pe}^2}{\omega(\omega +i\Gamma_e)} \vec{E} \qquad \vec{M} = -\frac{\mu_{0} \omega_{pe}^2}{\omega(\omega +i\Gamma_m)} \vec{H}. \label{polarizations}
\ee
Therefore, one obtains $\vec{D}$ and $\vec{B}$ fields in frequency domain from (\ref{Dfield}), (\ref{Bfield}) and (\ref{polarizations}) as follows
\begin{align}
\vec{D} (\vec{r}, \omega) &= \varepsilon_{0}\left[1-\frac{\omega_{pe}^2}{\omega(\omega +i\Gamma_e)} \right] \vec{E}(\vec{r}), \label{Dfield2}\\
\vec{B} (\vec{r}, \omega) &= \mu_{0}\left[1-\frac{\omega_{pm}^2}{\omega(\omega +i\Gamma_m)} \right] \vec{H}(\vec{r}). \label{Bfield2}
\end{align}
We demand to investigate the TZ mode configuration of this intriguing metamaterial environment. Here Z stands for (E)lectric or (M)agnetic field configurations\footnote{We use TZ = (TE, TM), but in field configurations Z denotes for E or H, i.e. $Z=(E, H)$.}, i.e. Z :=(E, M). For this purpose, we take $\vec{\nabla}\times$ of (\ref{curlE2}) for TE mode and (\ref{curlH2}) for TM mode to obtain the following 3-dimensional Helmholtz equation corresponding to TZ-mode
\begin{align}
\left[\nabla^2-\prod_{\ell=1}^{2}\left(ik-\frac{c k_{\ell}^{2}}{\Gamma_{\ell} -i\omega}\right)\right] \vec{Z}^{\alpha} (\vec{r})=0,\label{3dimHelm}
\end{align}
where $c:=1 /\sqrt{\varepsilon_{0}\mu_{0}}$, $k:=\omega /c$ and $k_{\ell}:=\omega_{p\ell} /c$ with $\ell = (1, 2) := (e, m)$. Here one can find the complementary field $\epsilon^{\alpha\beta}\vec{Z}^{\beta}$ from the Maxwell's equations in (\ref{curlH2}) and (\ref{curlE2}) 
\be
\epsilon^{\alpha\beta}\vec{Z}^{\beta} = (\zeta^{\beta})^{-1} \left[i\omega-\frac{\omega_{p\beta}^2}{\Gamma_{\beta}-i\omega}\right]^{-1} \vec{\nabla}\times\vec{Z}^{\alpha}. \label{Hfield3}
\ee

Here we define the parameter $\zeta := (\varepsilon_{0}, \mu_0)$ representing the generalized permittivity of the space for the field $\vec{Z}$ for convenience. Let the field $\vec{Z}$ be emitted along the $y$-direction in TZ-mode in the following form
\be
\vec{Z} (\vec{r}) = e^{ik_x x} Z_{y} (z)\,\hat{j}.\label{TE1}
\ee
Substituting (\ref{TE1}) in (\ref{3dimHelm}) gives rise to the following 1-dimensional Helmholtz equation
\be
Z_y^{''}+k_z^2\mathbf{\tilde{n}}^2 Z_y = 0,\label{1dimhelm}
\ee
where the symbol $\mathbf{\tilde{n}}$ can be interpreted as the complex-valued effective refractive index of the metamaterial environment in case of an oblique incidence angle $\theta$, $\mathbf{\tilde{n}} \in \mathbb{C}$ which is defined as follows
\begin{equation}
 \mathbf{\tilde{n}} := \frac{\sqrt{\mathbf{n}^2-\sin^2\theta}}{\cos\theta}.  \label{nt2} 
\end{equation} 
Here one can reveal the actual refractive index of the metamaterial medium by the following identification
\be
\mathbf{n} := \sqrt{\prod_{\ell=1}^{2}\left(1- \frac{\beta_{\ell}^2(1+\Gamma_{\ell} /i\omega)}{1+(\Gamma_{\ell} / \omega)^2}\right)}, \label{refractiveindex}
\ee
where $\beta_{\ell} := k_{\ell}/k$ is a dimensionless quantity. Notice that (\ref{1dimhelm}) is equivalent to the following 1-dimensional Schrödinger equation with a complex potential $v(\mathbf{z}):= k_z^2 (1-\mathbf{\tilde{n}}^2)$, given by the piece-wise continuous map $v(\mathbf{z}):\mathbb{C}\rightarrow\mathbb{C}$ with a compact support, and energy $E := k_z^2$,
\be
-\psi''(\mathbf{z}) + v(\mathbf{z})\psi (\mathbf{z}) = E \psi (\mathbf{z}).
\ee
Solution of the Helmholtz equation in (\ref{1dimhelm}) gives rise the following field expressions of $\vec{E}$ and $\vec{H}$ respectively,
\begin{align}
\vec{Z}^{\alpha} (x, z) &= e^{ik_x x}\mathcal{F}_{+}(z)\hat{j},\notag\\
\epsilon^{\alpha\beta}\vec{Z}^{\beta} (x, z) &= \frac{ike^{ik_x x}}{\zeta\left[i\omega-\frac{\omega_{p\beta}^2}{\Gamma_{\beta}-i\omega}\right]}\left[\mathcal{F}_{+}(z)\sin\theta \hat{k}-\mathbf{\tilde{N}}\mathcal{F}_{-}(z)\cos\theta \hat{i}\right],\notag
\end{align}
where we denote the following specifications in different regions of space for convenience
\begin{align}
	\mathcal{F}_{\pm}(z) : = \begin{cases}
		A_1 e^{ik_z z} \pm  B_1 e^{-ik_z z},\,\,\,\,\,\,\,\,\,\,\,\,z < 0\\
		A_2 e^{ik_z\mathbf{\tilde{n}}z} \pm B_2 e^{-ik_z\mathbf{\tilde{n}}z},\,\,\,\,\,\,0 < z < L\\
		A_3 e^{ik_zz} \pm  B_3 e^{-ik_zz},\,\,\,\,\,\,\,\,\,\,\,\,z > L.
	\end{cases} \label{Fpm}
\end{align}
\begin{align}\label{Nt2}
	\mathbf{\tilde{N}} : = \begin{cases}
		1,\,\,\,\,\,\,\,\,\,\,\,\,z < 0,\\
		\mathbf{\tilde{n}},\,\,\,\,\,\,0 < z < L,\\
		1,\,\,\,\,\,\,\,\,\,\,\,\,z > L.
	\end{cases} 
\end{align}
Furthermore, one can obtain $\vec{D}$ and $\vec{B}$ fields in all regions of space by employing (\ref{Dfield2}) and (\ref{Bfield2}),
\begin{align}
\vec{U}^{\alpha} (x, z) &= \zeta\left[1-\frac{\omega_{p\alpha}^2}{\omega(\omega +i\Gamma_{\alpha})} \right] e^{ik_x x}\mathcal{F}_{+}(z)\hat{j},\notag\\
\epsilon^{\alpha\beta}\vec{U}^{\beta} (x, z) &= \frac{1}{c}e^{ik_x x}\left[\mathcal{F}_{+}(z)\sin\theta \hat{k}-\mathbf{\tilde{N}}\mathcal{F}_{-}(z)\cos\theta \hat{i}\right],\notag
\end{align}
where we defined $c:=\omega / k$, and the field $\vec{U}$ stands for the $\vec{D}$ and $\vec{B}$ fields, i.e. $\vec{U} := (\vec{D}, \vec{B})$. Having obtained all field profiles, which are provided in Table~(\ref{t1}), one can apply the relevant boundary conditions, which are specified as follows. Let $\mathcal{S}$ be an interface between two regions of space on which there is a surface charge density $\rho^s$ and current density $\vec{\cJ}^s$. Let $\hat{n}$ be the unit normal vector to the surface $\mathcal{S}$ extending  from region 2 to region 1.  Boundary conditions across the surface $\mathcal{S}$ between two regions of space are given by the statements: 1) Tangential component of electric field $\vec{E}$ is continuous across the interface, $\hat{n}\times (\vec{E}_1-\vec{E}_2) = 0$; 2) Normal component of magnetic field vector $\vec{B}$ is continuous, $\hat{n}\cdot (\vec{B}_1-\vec{B}_2) = 0$; 3) Normal component of electric flux density vector $\vec{D}$ is discontinuous by an amount equal to the surface current density, $\hat{n}\cdot (\vec{D}_1-\vec{D}_2) =\rho^s$; 4) Tangential component of the field $\vec{H}$ is discontinuous by an amount equal to the surface current density, $\hat{n}\times (\vec{H}_1-\vec{H}_2) = \vec{\cJ}^s$. In our wave configuration, the third condition does not exist since there is no normal component of $\vec{D}$ field. Also, surface current density can be written in components by $\vec{\cJ}^s_{\alpha} = \sigma_{\alpha\beta} E_{\beta}$\footnote{Here, indices $\alpha$ and $\beta$ may cause a confusion. They are not related to TE or TM mode symbols, but they represent components of the associated currents.}, where $\sigma$ is the conductivity on the surface. Therefore, the boundary conditions arise as in Table~\ref{table1}. 
\begin{table*}[t]
	\centering
	\scalebox{1.10}{\begin{tabular}{|c|c|c|c|c|}
				\hline
					&$\vec{E} (z) e^{ik_xx}$& $\vec{D}(z) e^{ik_xx}$& $\vec{B}(z) e^{ik_xx}$& $\vec{H}(z) e^{ik_xx}$\\
				\hline\hline
					&$E_x = 0$ & $D_x = 0$ &  $B_x = - c^{-1}\mathbf{\tilde{N}}\mathcal{F}_{-}(z)\cos\theta$ & $H_x = -ik\mu_0^{-1}\mathbf{a}_{m}^{-1}\mathbf{\tilde{N}}\mathcal{F}_{-}(z)\cos\theta$\\
				 TE&\raggedleft $E_y = \mathcal{F}_+(z)$ & $D_y = -i\varepsilon_{0} \omega^{-1}\mathbf{a}_{e} \mathcal{F}_{+}(z)$ & $B_y = 0$ & $H_y = 0$\\
				&$E_z = 0$ & $D_z = 0$ & $B_z = c^{-1}\mathcal{F}_{+}(z)\sin\theta$ & $H_z = ik\mu_0^{-1}\mathbf{a}_{m}^{-1}\mathcal{F}_{+}(z)\sin\theta$\\ 
				\hline
    				&$E_x = ik\epsilon_0^{-1}\mathbf{a}_{e}^{-1}\mathbf{\tilde{N}}\mathcal{F}_{-}(z)\cos\theta$ & $D_x = c^{-1}\cos\theta\mathbf{\tilde{N}}\mathcal{F}_-(z)$ &  $B_x = 0$ & $H_x = 0$\\
				TM &\raggedleft $E_y = 0$ & $D_y = 0$ & $B_y = -i\mu_{0}\omega^{-1} \mathbf{a}_{m} \mathcal{F}_{+}(z)$& $H_y = \mathcal{F}_+(z)$\\
				&$E_z = -ik\epsilon_0^{-1}\mathbf{a}_{e}^{-1}\mathcal{F}_{+}(z)\sin\theta$ & $D_z = -c^{-1} \sin\theta\mathcal{F}_+(z)$ & $B_z = 0$ & $H_z = 0$\\ 
				\hline
			\end{tabular}}	
		\caption{ Components of the $\vec{E}$, $\vec{D}$, $\vec{B}$, $\vec{H}$ fields inside and outside of Metamaterial slab corresponding to TE and TM modes are shown in the table respectively. Here quantities $\mathcal{F}_{\pm}(z)$ are specified in \ref{Fpm}, the effective refractive index of whole space is identified in \ref{Nt2} and  $\mathbf{a}_{\ell} := i\omega-\frac{\omega_{p\ell}^2}{\Gamma_{\ell}-i\omega}$.} \label{t1}
\end{table*}
\begin{table*}[ht]
	\centering
	\scalebox{1.25}{\begin{tabular}{|l|}
				\hline
				\quad\quad\quad\quad\quad\quad\textbf{Boundary Conditions}  \\
				\hline\hline
					$  A_1 +  B_1 = A_2 + B_2 $  \\
				$  A_2 e^{ik_z\mathbf{\tilde{n}}L} + B_2 e^{-ik_z\mathbf{\tilde{n}}L} = A_3 e^{ik_zL} + B_3 e^{-ik_zL}$ \\
				$  \Bar{\mathbf{n}}_{\ell}\left[B_2 -A_2\right] = \Bar{\sigma}_{1-} A_1 + \Bar{\sigma}_{1+}B_2$\\ 
    $  \Bar{\mathbf{n}}_{\ell}\left[A_2 e^{ik_z\tilde{\mathbf{n}}L} - B_2 e^{-ik_z\tilde{\mathbf{n}}L}\right] =  \Bar{\sigma}_{2+} A_3 e^{ik_zL} + \Bar{\sigma}_{2-} B_3 e^{-ik_zL}$\\
    \hline
    \end{tabular}}
    \caption{Table depicts the boundary conditions for the metamaterial slab adjusted in the TE mode. Here, we identify the quantities $\Bar{\mathbf{n}}_{\ell}$ and $\Bar{\sigma}_{j\pm}$ in (\ref{identification1}). $\ell = (e, m)$ corresponds to the TE and TM mode configurations. }\label{table1}
\end{table*}
In this table, we identify the following shorthand notations for convenience
\be
\Bar{\mathbf{n}}_{\ell}:= \frac{Z_0 \mathbf{\tilde{n}}\cos\theta}{1-\frac{\omega_{p\ell}^2}{\omega(\omega +i\Gamma_{\ell})}}\qquad \Bar{\sigma}_{j\pm} := \sigma_{j, x y} \pm Z_0 \cos\theta.\label{identification1}
\ee
Here, $\sigma_{j, x y}$ implies the surface conductivity of the $j^{th}$ surface in tensorial form, and $\ell = (e, m)$ corresponding to TE and TM modes. This is the only conductivity component that can survive on the surface, which implies that there should be a Kerr/Faraday rotations when the wave hits on the surface. But we know that Kerr/Faraday rotations do not arise unless system contains an axion term, which is not the case in our configuration. Therefore, an induced current can not form on the surface of the metamaterial unless we have free surface currents. Hence, with the help of these boundary conditions, it is possible to construct the transfer matrix. As is well-known, Transfer matrix is one of the most important tools that contains all the significant information about the scattering system. It joins incoming waves to the outgoing waves in such a way that all reflection and transmission amplitudes are contained within the matrix. Once we define the incoming and outgoing states as follows
\begin{align}
&\bC_{k} := \left(
                       \begin{array}{c}
                         A_k \\
                         B_k \\
                       \end{array}
                     \right),\notag
\end{align}
with $k=(1, 3)$, one can incorporate them by means of the transfer matrix $\mathbb{M}$ as follows
\begin{equation}
\bC_{3} = \mathbb{M}\,\bC_{1}.
\end{equation}

\begin{figure}
    \centering
    \includegraphics[width=8 cm]{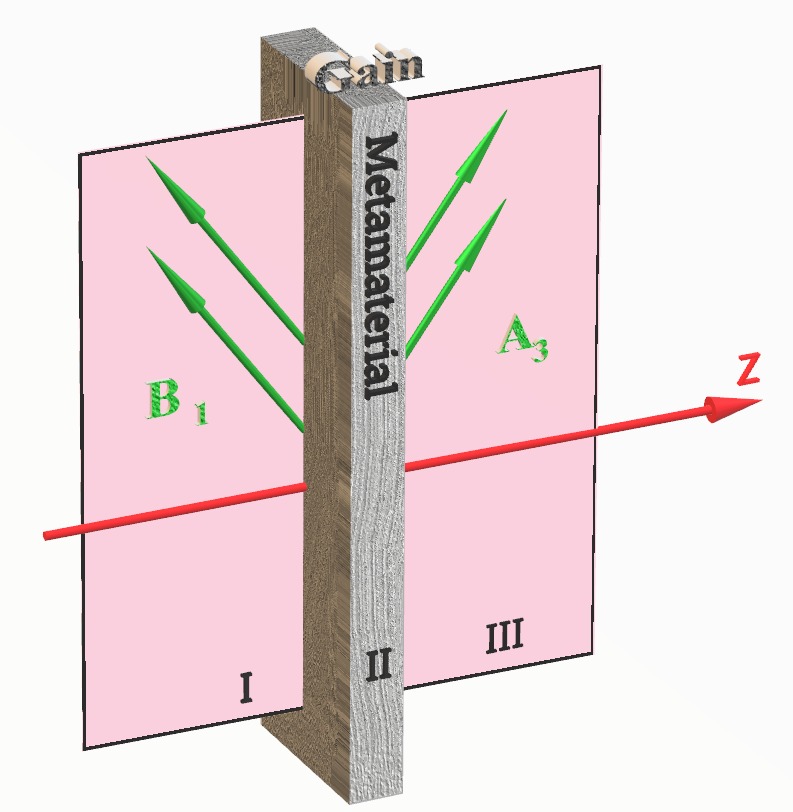}
    \caption{Perspective view of lasing configuration in a metamaterial slab, arising from spectral singularity condition in (\ref{spectralsing}).}
    \label{figlase}
\end{figure}

Spectral singularities correspond to the real $k$-values which cause purely outgoing waves, see Fig.~(\ref{figlase}), resulting in divergent reflection and transmission amplitudes in the metamaterial system\cite{naimark, p123}. It is in fact leading to zero width resonance condition, see\cite{CPA, lastpaper} for a list of literature on this effect. In our case, it corresponds to the real zeros of $M_{22}$ component of the transfer matrix $\mathbb{M}$, i.e. $M_{22} (k) = 0$. Therefore, the spectral singularity condition for TZ mode is obtained as follows
\be
e^{2i k_z \mathbf{\tilde{n}} L} = \frac{(\Bar{\mathbf{n}}_{\ell} + \Bar{\sigma}_{2+})(\Bar{\mathbf{n}}_{\ell} + \Bar{\sigma}_{1+})}{(\Bar{\mathbf{n}}_{\ell} - \Bar{\sigma}_{2+})(\Bar{\mathbf{n}}_{\ell} - \Bar{\sigma}_{1+})},\label{spectralsing}
\ee
where $\ell = (e, m)$ correspond to TE and TM mode, respectively. This equation produces the optimal parameters of the metamaterial system for the lasing threshold condition. 

Now, we get back to the refractive index given by (\ref{refractiveindex}). We realize that refractive index can be written in terms of real and imaginary part as follows
\be
\mathbf{n} := \eta + i \kappa . \label{refindex2}
\ee
To find $\eta$ and $\kappa$ using (\ref{refractiveindex}), one obtains the following expressions
\begin{align}
2\eta\kappa = &\frac{\beta_e \Gamma_e/\omega}{1 + (\Gamma_e/\omega)^2} + \frac{\beta_m \Gamma_m/\omega}{1 + (\Gamma_m/\omega)^2}\notag\\ 
&+ \frac{(\beta_e\beta_m/\omega) (\Gamma_e + \Gamma_m)}{[1 + (\Gamma_e/\omega)^2][1 + (\Gamma_m/\omega)^2]},\label{realref}\\
\eta^2-\kappa^2 = &1-\frac{\beta_e}{1 + (\Gamma_e/\omega)^2}-\frac{\beta_m}{1 + (\Gamma_m/\omega)^2}\notag\\
&+ \frac{\beta_e\beta_m [1-\Gamma_e\Gamma_m/\omega^2] }{[1 + (\Gamma_e/\omega)^2][1 + (\Gamma_m/\omega)^2]}.\label{imaginaryref}
\end{align}
We denote the magnitude and phase of $\mathbf{n}$ by $|\mathbf{n}|$ and $\vartheta$, respectively, such that they are found out to be
\begin{align}
|\mathbf{n}| &= \sqrt{\eta^2 + \kappa^2}, \notag\\
\vartheta &= \frac{1}{2}\tan^{-1} \left(\frac{2\eta\kappa}{\eta^2-\kappa^2}\right).
\end{align}
In practice, $|\mathbf{n}|$ is always positive, and depending on the phase angle $\vartheta$, refractive index $\mathbf{n}$ may admit a negative values in both real and imaginary parts. For its real part to be negative definite regardless of the imaginary part, one must impose the condition
\be
\frac{\pi}{2} < \vartheta <\frac{3\pi}{2}.
\ee
But if we desire both real and imaginary parts to have negative values, then phase angle is restricted to the range
\be
\pi < \vartheta <\frac{3\pi}{2}\quad\Longrightarrow\quad 2\pi < \tan^{-1} \left(\frac{2\eta\kappa}{\eta^2-\kappa^2}\right)< 3\pi .\label{negrefind2}
\ee
This in fact determines the lasing threshold condition for a negative index metamaterial. Similarly, we can consider a negative real part in conjunction with a positive imaginary part, which gives rise to the condition for coherent perfect absorption. This occurs only if
\be
\frac{\pi}{2} < \vartheta <\pi\quad\Longrightarrow\quad \pi < \tan^{-1} \left(\frac{2\eta\kappa}{\eta^2-\kappa^2}\right)< 2\pi .\label{negrefind3}
\ee

Next, for our present purposes, we wish to compute the negative refractive index condition given in (\ref{negrefind2}) depending on the parameters of the medium. We recall the identity
\be
\tan^{-1} (z) = \pi m +\frac{1}{2i}\ln \left(\frac{1+iz}{1-iz}\right).
\ee
where $z:= \frac{2\eta\kappa}{\eta^2-\kappa^2}$ is an arbitrary real parameter and $m$ is an integer. Thus, Eq.~((\ref{negrefind2})) turns into 
\be
0 < (m-2) \pi -i\ln \left(\frac{\mathbf{n}}{\mathbf{n}^{*}} \right) < \pi ,
\ee
which is reduced to the form
\be
(2-m) \pi\,> \,2\tan^{-1}\left(\kappa/\eta \right)\,>\,(3-m) \pi . \label{cond1}
\ee
Recall that, for most materials with positive refractive index, the condition $|\kappa|<<\eta$ holds safely. In our case of metamaterials with a possible flexible (negative) refractive indices, we may adopt the condition $|\kappa|<<|\eta|$, which is typical for many materials. However, it should also be noted that due to the highly dispersive nature of metamaterials at optical frequencies, they exhibit significant loss. This makes it challenging to develop applications that depend on gain media. Nevertheless, we will conduct our analysis on the assumption that these practical challenges have been successfully overcome. Therefore, $\tan^{-1}\left(\kappa/\eta \right) \approx \kappa/\eta$. Thus, (\ref{cond1}) yields the required range of lasing threshold condition arising from a metamaterial slab, which is expressed explicitly in terms of system parameters as follows
\begin{widetext}
\begin{equation}
(2-m) \pi\,< \,\frac{(\beta_e\Gamma_e/\omega) [1 +\beta_m + (\Gamma_m/\omega)^2] + (\beta_m\Gamma_m/\omega) [1 +\beta_e + (\Gamma_e/\omega)^2]}
{(\beta_e\Gamma_e\Gamma_m/\omega^2) - [1 - \beta_m + (\Gamma_m/\omega)^2][1 - \beta_e + (\Gamma_e/\omega)^2] }  \,<\,(3-m) \pi . 
\end{equation} 
\end{widetext}
Notice that $m$ acts as a winding parameter. The imaginary part of the refractive index determines the gain coefficient of the medium,
\be
    g := -2\kappa k = -\frac{4\pi \kappa}{\lambda}.\label{gainn}
\ee
Gain in metamaterials refers to the process in which electromagnetic waves are strengthened as they pass through these specially designed materials. Metamaterials can either boost signal strength (gain) or weaken it (loss), which is useful for things like amplifiers or sensors \cite{gainmat1, gainmat2, gainmat3, gainmat4}. Finding the right balance between gain and loss is important for controlling wave behavior and enabling advanced technologies in fields like communication and photonics.

We can perturbatively analyze the condition and see the influence of the associated parameters of the system. Employing the condition $|\kappa| \ll |\eta|$ in the definition of $\mathbf{\tilde{n}}$ given by (\ref{nt2}), and substituting in definitions of $\mathbf{\tilde{n}}$ in (\ref{nt2}), one obtains the following results for the gain coefficient $g_{\ell}$ and $k_{\ell}$ up to the order of $\mathcal{O} (\kappa^2)$,
\begin{align}\label{eq27}
    g_{\ell} &=\frac{\tilde{\eta}\cos\theta}{\eta L} \ln\left|\frac{(\Bar{\mathbf{n}}_{\ell} + \Bar{\sigma}_{2+})(\Bar{\mathbf{n}}_{\ell} + \Bar{\sigma}_{1+})}{(\Bar{\mathbf{n}}_{\ell} - \Bar{\sigma}_{2+})(\Bar{\mathbf{n}}_{\ell} - \Bar{\sigma}_{1+})}\right|,\notag\\
    k_{\ell} &= \frac{1}{L\cos\theta\tilde{\eta}}\left\{ \tan^{-1}\varphi_{\ell}+ m\pi \right\},
\end{align}
where $\ell$ denotes the TE and TM modes, $\varphi_{\ell}:= \arg(\mathbf{z}_{\ell})$ stands for the principle argument (phase angle) of relevant expression of $\mathbf{z}_{\ell} = \left[\frac{(\Bar{\mathbf{n}}_{\ell} + \Bar{\sigma}_{2+})(\Bar{\mathbf{n}}_{\ell} + \Bar{\sigma}_{1+})}{(\Bar{\mathbf{n}}_{\ell} - \Bar{\sigma}_{2+})(\Bar{\mathbf{n}}_{\ell} - \Bar{\sigma}_{1+})}\right]$, $m$ represents integer values, i.e. $m= 0, \pm 1, \pm 2, \dots$, and specifications $\tilde{\eta}$ and $\tilde{\kappa}$ are respectively the effective real and imaginary parts of the material environment in case of oblique incidence, i.e. $\mathbf{\tilde{n}} \approx \tilde{\eta} + i \tilde{\kappa}$, which are defined as below,
\begin{align}
    \tilde{\eta} :&= \frac{\sqrt{\eta^2-\sin^2\theta}}{\cos\theta},\\
    \tilde{\kappa}:&=\frac{\eta \kappa}{\tilde{\eta} \cos^2\theta}.
\end{align}

\section{B. Refractive Index and Spectral Singularities in TE and TM Modes}

We observe that to obtain a laser from a metamaterial slab, the refractive index condition given by Eq.~(34) must first be met. Subsequently, the spectral singularity condition given by Eq.~(\ref{spectralsing}) must be fulfilled.

\begin{figure}
    \centering
    \includegraphics[width=8.5cm]{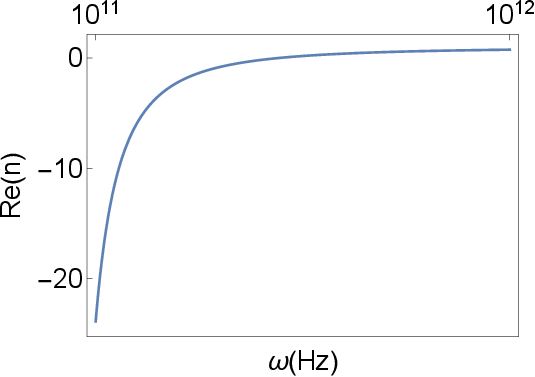}\\
    \includegraphics[width=8.5cm]{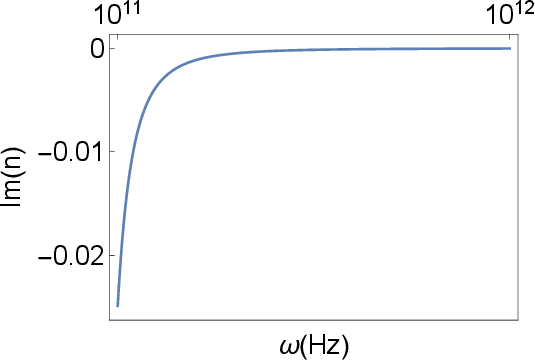}
    \caption{Graphs showing the real and imaginary parts of refractive index given by (\ref{refindexw}). We use the specifications in (\ref{specification}). Notice that both real and imaginary parts in designated range are negative.}
    \label{fign}
\end{figure}

\begin{figure}
    \centering
    \includegraphics[width=8.5cm]{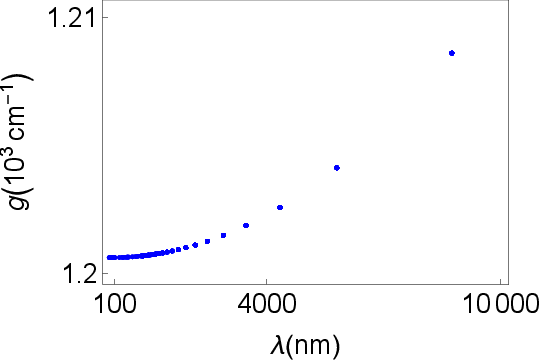}
    \caption{Pictorial figure showing the spectral singularity points in $\lambda - g$ plot for both TE and TM mode. We use the specifications in (\ref{specification}). Angle of incidence is chosen to be $\theta=30^{\circ}$ and the thickness of the slab is $L= 10~\mu m$.}
    \label{figspect}
\end{figure}

To find out the physical parameters of the system corresponding to the spectral singularity condition, let's consider a metamaterial whose characteristics are provided as follows\cite{Li2008, Ziolkowski2003}
\begin{align}
&\Gamma_e = \Gamma_m = \Gamma= 10^8~\textrm{s}^{-1},~~  \omega_{pe} = \omega_{pm}= \omega_{p}= 2\pi\sqrt{7} f_0,\notag\\ &\mathbf{n} (\omega_0)\approx -6, ~~ \omega_0= 2\pi f_0,~~f_0= 30~\textrm{GHz}, ~~\lambda_0 = 10^{-2}~\textrm{m}. \label{specification}
\end{align}
Here $f_0$ is the center frequency for the index of refraction corresponding to the wavelength of the free space $\omega_0$. Refractive index in (\ref{refractiveindex}) depending upon frequency $\omega$ with these specifications turns out to be
\be
\mathbf{n} (\omega) = 1 - \frac{\omega_p^2}{\omega(\omega - i\Gamma)}. \label{refindexw}
\ee  
Fig.~(\ref{fign}) shows the possible values that the real and imaginary parts of the refractive index can take. As can be seen, the real and imaginary parts are negative in the selected frequency ranges as desired. Thus, it is evident at which frequency values the refractive index can exhibit negative values, depending on the intended purpose. In our case, since we aim to keep the imaginary part of the refractive index as small as possible, it is most suitable to use higher frequency values whenever feasible (i.e., frequencies near $10^12$ Hz). We realize that properties of the metamaterial slab provided by ((\ref{specification})) give rise to overlapping TE and TM modes (TEM mode). This is a special case that causes spectral singularity points to coincide, see Fig.~(\ref{figspect}).

In Fig.~(\ref{figspect}), the placement of spectral singularity points can be seen in $\lambda - g$ plane. These points are the same for both TE and TM mode. As can be clearly understood, as the wavelength value increases, the frequency of points where spectral singularity values can be taken decreases. Gain values at higher wavelengths increase very slightly. In order to obtain a laser from a metamaterial, it is necessary to use the system parameters specified in these points.

\begin{figure}
    \centering
    \includegraphics[width=8.5cm]{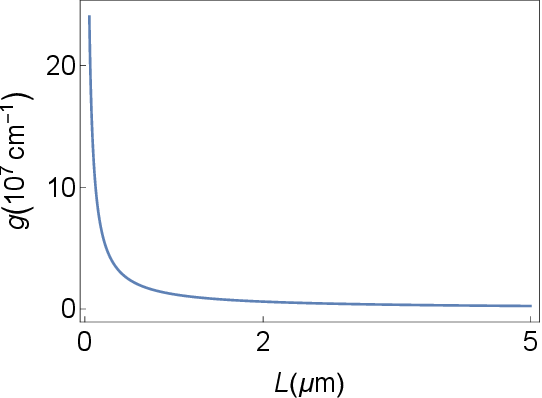}
    \caption{Figure shows the dependence of gain value upon thickness of the slab $L$. Angle of incidence is opted to be $\theta=30^{\circ}$. Increasing the thickness of the slab leads to decrease gain value required for lasing significantly.}
    \label{figgl}
\end{figure}

Fig.~(\ref{figgl}) shows how the gain value $g$ changes with the thickness of the slab $L$. As the thickness of the slab decreases, the corresponding gain value increases significantly. Therefore, choosing relatively larger slab thicknesses considerably reduces the gain value.

\begin{figure*}
    \centering
    \includegraphics[width=8.5cm]{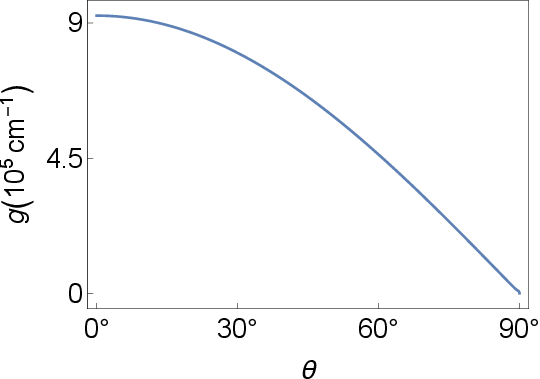}
    \includegraphics[width=8.2cm]{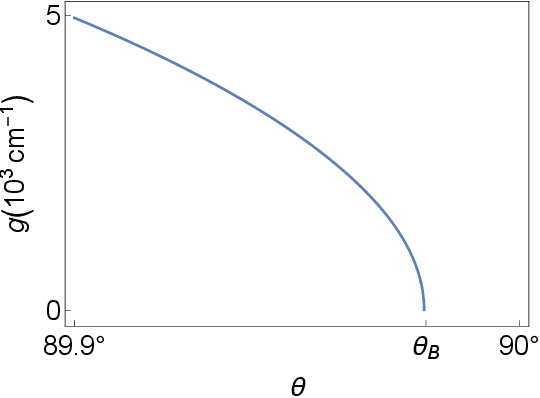}
    \caption{Figure depicts the dependence of gain value upon the angle of incidence $\theta$. Right panel cover the whole range while the left panel clearly indicates the Brewster's angle around $90^{\circ}$. Here thickness of the slab is $L= 1.5~\mu m$. As the angle of incidence increases, the corresponding gain value decreases considerably. Notice that Brewster's angle is at $\theta_B \approx 89.979^{\circ}$.}
    \label{figgangle}
\end{figure*}

Fig.~(\ref{figgangle}) demonstrates how the gain value $g$ varies with the incidence angle $\theta$. The most suitable angles are around the angles further from the surface normal. However, it will be seen that the Brewster's angle has an effect here. This angle determines the maximum incidence angle value that the system can achieve in lasing, even if it is very close to 90 degrees for the material properties that we use. In terms of Brewster's angle, the system does not lase at high angles. To find the Brewster's angle in our case, one computes
\be
\left|\frac{\mathbf{\tilde{n}}}{1-\frac{\omega_{p\ell}^2}{\omega(\omega +i\Gamma_{\ell})}}-1\right| = 0.\label{brewster}
\ee
which leads to $\theta_B \approx 89.979^{\circ}$ for the values given in (\ref{specification}). This observation has not been explored in the existing literature. For studies on the impact of the incidence angle on metamaterial absorption, refer to \cite{absorp1, absorp2}.

On the other hand, if the metamaterial of interest possesses distinct plasma and damping frequencies corresponding to TE and TM modes, then we may observe the resolution of these modes. In this case refractive index will not have the form of Eq.~(\ref{refindexw}), but we must use the original expression in Eq.~(\ref{refractiveindex}). In this case, one can compute real and imaginary parts of refractive index using Eqs.~(\ref{realref}) and (\ref{imaginaryref}) as follows,
\begin{align*}
&\eta(\omega) \approx \sqrt{1-\frac{\omega\omega_{pe}}{\omega^2 + \Gamma_e^2}-\frac{\omega\omega_{pm}}{\omega^2 + \Gamma_m^2}
+ \frac{\omega_{pe}\omega_{pm} [\omega^2-\Gamma_e\Gamma_m] }{[\omega^2 + \Gamma_e^2][\omega^2 + \Gamma_m^2]}},\notag\\
&\kappa(\omega) \approx \frac{\frac{\omega_{pe} \Gamma_e}{\omega^2 + \Gamma_e^2} + \frac{\omega_{pm} \Gamma_m}{\omega^2 + \Gamma_m^2}
+ \frac{(\omega_{pe}\omega_{pm} \omega) (\Gamma_e + \Gamma_m)}{[\omega^2 + \Gamma_e^2][\omega^2 + \Gamma_m^2]}}{2\sqrt{1-\frac{\omega\omega_{pe}}{\omega^2 + \Gamma_e^2}-\frac{\omega\omega_{pm}}{\omega^2 + \Gamma_m^2}
+ \frac{\omega_{pe}\omega_{pm} [\omega^2-\Gamma_e\Gamma_m] }{[\omega^2 + \Gamma_e^2][\omega^2 + \Gamma_m^2]}}}.\notag
\end{align*}

\begin{figure*}
    \centering
    \includegraphics[width=8.4cm]{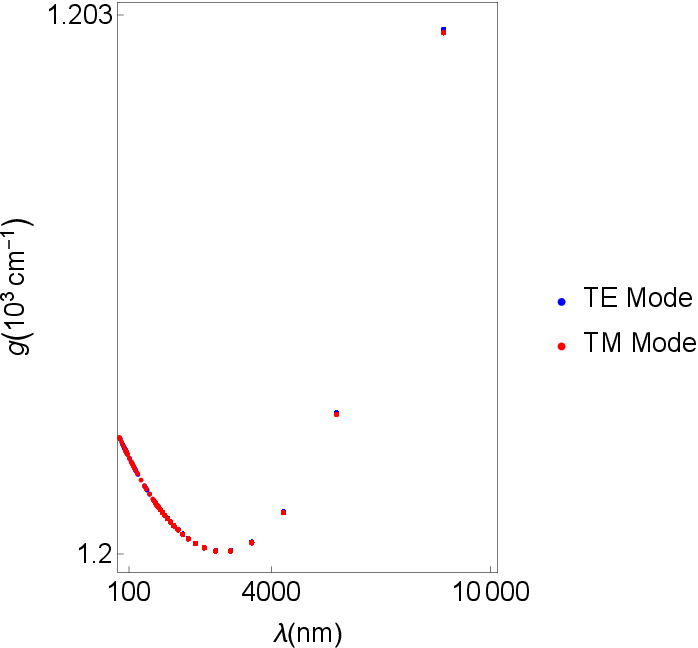}
    \includegraphics[width=8.8cm]{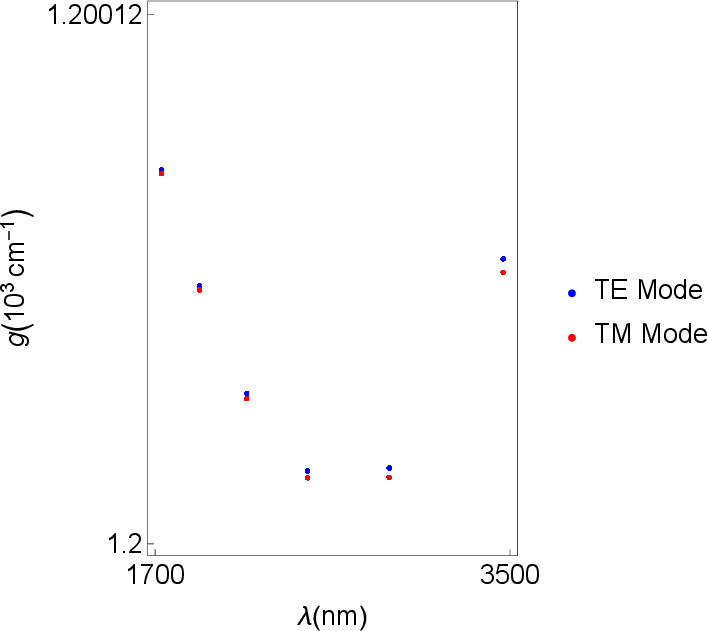}
    \caption{Figure shows the distribution of the spectral singularities in gain $g$ and the wavelength $\lambda$ plane for different plasma and damping frequencies. Our selected frequency values are $\omega_{pe}=10^{12}~s^{-1}$, $\omega_{pm}=10^{5}~s^{-1}$, $\Gamma_{e}=10^{8}~s^{-1}$ and $\Gamma_{m}=10^{4}~s^{-1}$. We observe that TE and TM modes distinguish from each other as the wavelength increases.}
    \label{figglamda2}
\end{figure*}

One may find the resonance frequency $\omega_0$ at resonant wavelength $\lambda_0$ such that $\omega_0 = 2\pi c /\lambda_0$. Thus, the refractive index $\mathbf{n}_0$ at resonance case can be determined as $\mathbf{n}_0 = \eta_0 + i \kappa_0$, where $\eta_0= \eta(\omega_0)$ and $\kappa_0=\kappa(\omega_0)$. For instance, as a demonstration, we consider a hypothetical metamaterial sample possessing the special material properties as $\omega_{pe}=10^{12}~s^{-1}$, $\omega_{pm}=10^{5}~s^{-1}$, $\Gamma_{e}=10^{8}~s^{-1}$ and $\Gamma_{m}=10^{4}~s^{-1}$ together with the designed incidence angle $\theta = 30^{\circ}$ and slab thickness $L=1.5~\mu m$, one gets the configuration in Fig.~(\ref{figglamda2}). It is obvious that TE and TM modes resolve from each other. As the wavelength increases, the resolution becomes more prominent. Therefore, lasing modes in TE and TM cases are separated from each other at almost the same wavelength. As can be clearly seen, it is easier for the NIM system to lase in TM mode at the given values. 

\section{C. Construct of CPA in Metamaterials}

\begin{figure}
    \centering
    \includegraphics[width=8 cm]{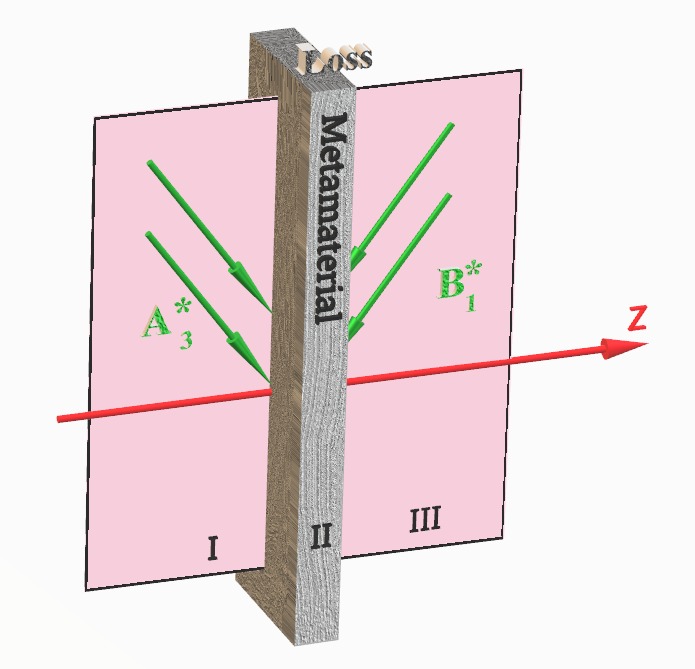}
    \caption{Perspective view of CPA configuration in a metamaterial slab, arising from spectral singularity condition in (\ref{spectralsing}), but now gain $\mathbf{n}$ is replaced by loss $\mathbf{n}^{*}$.}
    \label{figcpa}
\end{figure}

Another interesting application of an active (NIM) metamaterial slab is the coherent perfect absorption (CPA)\cite{fu2017, chen2018, monks2018, ji2021, hu2016, lin1, lin2}. It is already known that CPA occurs at the same spectral singularity points such that system acts as a time-reversed system\cite{CPA, CPA2}. In this case, one just needs to switch the gain media into the lossy one by just taking the complex conjugate of $\mathbf{n}$, $\mathbf{n} \rightarrow \mathbf{n}^{*}$, such that metamaterial system starts to absorb electromagnetic waves perfectly at the same parameters as a lasing system. This means that CPA configuration leads to the same figures that we found above for a lossy environment.  In this case slab system will produce the following configuration in Fig.~(\ref{figcpa}),

\section{Concluding Remarks}

In this study, we obtained the necessary conditions for the construction of highly effective (NIM) metamaterial lasers with the help of non-Hermitian physics. The results we found are quite striking and will guide studies in this field. To this end, we examined the existence of spectral singularities resulting from the interaction of a NIM system with time harmonic electromagnetic waves. Since the interaction of such materials with electromagnetic waves is rather incompatible from the usual materials, we solved the relevant governing equations and obtained how the refractive index and boundary conditions of the material are shaped depending on the plasma and damping frequencies. Ultimately, by obtaining the transfer matrix, we pointed out how the parameters of the system can be employed to obtain specific configurations of the system. We are able to extract the spectral singularity condition for a (NIM)metamaterial laser system in (\ref{spectralsing}). This is the lasing threshold condition for a NIM laser system.

While the spectral singularity condition is mathematically defined in our study, its experimental realization presents significant challenges. Effective loss management is critical for practical implementations, especially in non-Hermitian systems, and addressing fabrication constraints and target wavelength ranges will enhance the study's relevance to real-world applications. In fact, the target wavelength range for metamaterials spans from microwaves to optical frequencies, with fabrication challenges increasing at higher frequencies due to the need for precise nanoscale features. In non-Hermitian systems, effective loss control is crucial, as unmitigated losses can degrade performance, making careful material selection and structural optimization essential for practical implementations\cite{ct1}.

Notice that the large fabrication cost of the metamaterial slabs can make them unfavourable in practical uses like lasers discussed here. However, to make this attractive, it is necessary to make laser production from such materials feasible in terms of appropriate parameters. Since the approach we use calculates exactly the lasing threshold condition, this problem is eliminated. In the light of our results, we determined the ranges of lasing parameters of such material types by obtaining the condition of having a negative imaginary part while maintaining the negative value of the refractive index. This is crucial because normally not all material properties may lead to lasing together with sustaining metamaterial properties. Our findings prove that (NIM) metamaterial system lases as long as the conditions (34) and (\ref{spectralsing}) hold. Besides, electric and magnetic plasma and damping frequencies determine TE and TM mode spectral singularity points. If they get the same values, then we observe that both modes coincide and we just observe a single mode. Moreover, once these features are distinguished considerably, TE and TM modes are resolved from each other. The interfaces between metamaterials with distinct plasma frequencies can lead to unique electromagnetic behaviors, such as changes in reflection, transmission, and absorption. These interactions can enable novel applications in wave manipulation, including sensing, cloaking, and imaging\cite{plasma}.

Another interesting feature is the nonlinear effects in metamaterials. Nonlinear effects in metamaterials arise when the material's response to electromagnetic fields becomes dependent on the intensity of the incident wave, leading to phenomena such as second-harmonic generation, self-focusing, and bistability. These effects are particularly significant in metamaterials with engineered structures that can exhibit strong non-linearities even at low input powers. Nonlinear metamaterials have promising applications in areas like optical switching, signal processing, and enhanced sensing due to their ability to manipulate light in novel ways. We will clearly reveal the effects on this issue in our future studies.

Furthermore, we explore that both TE and TM modes may lead to appear Brewster's angle, which is given by the condition in Eq.~(\ref{brewster}). Since metamaterials constitute ideal artificial material types for purpose-oriented material use, there can be very interesting application areas, especially in the fields of invisibility, CPA and laser physics. In particular, invisibility applications are among the popular application areas of such materials. The results we found for this can be used to construct a PT-symmetric system. We know that the invisibility phenomenon can be achieved with a PT-symmetric systems readily. This will be one of our potential future working topics. The method we used and the results we found in this study are very important and have the potential to provide groundbreaking facilities for metamaterial studies. Considering the active use of metamaterials in many areas of technology in the near future, the effective results of our work will be better understood.

Finally, when the input power is sufficiently larger than the lasing threshold condisiton in NIM systems we discussed here, multimode lasing takes place. Thus, non-linearity like the gain saturation plays an important role. Since this situation is not considered in our study, only a linear environment and a weak signal case are considered. However, the existence of nonlinearity in NIM systems can produce quite interesting results by considering that the system parameters can be used as a tuning parameter. This situation needs to be examined separately. However, for the purposes of this study, we do not include this analysis here.

\section{Acknowledgement}

We express our gratitude to the referees for their valuable contributions and insightful feedback during the review process of this paper.

\section{Remarks}

\textbf{Contribution Statement:} All authors contributed equally to the whole work including calculations, analysis and manuscript typing.

\textbf{Data Availability Statement:} Data supporting this study are included within the article and/or supporting materials.

\textbf{Competing Interests:} The authors declare no competing interests.


\begin{thebibliography}{99}



\bibitem{veselago} V.~G.~Veselago, ``The electrodynamics of substances with simultaneously negative permittivity and permeability," Soviet physics solid state, \textbf{8(11)} , 2500-2501, (1967).

\bibitem{pendry2000} J.~B.~ Pendry, ``Negative refraction makes a perfect lens," Physical review letters, \textbf{85(18)}, 3961, (2000).

\bibitem{eleftheriades2005} G.~V.~Eleftheriades and K.~G~.Balmain. ``Negative-refraction metamaterials: fundamental principles and applications". John Wiley and Sons, (2005).

\bibitem{agranovich2004} V.~M.~Agranovich, et al. ``Linear and nonlinear wave propagation in negative refraction metamaterials," Physical Review B, \textbf{69.16}, 165112, (2004).

\bibitem{smith2004} D.~R.~Smith, J.~B.~Pendry, and Mike CK Wiltshire, ``Metamaterials and negative refractive index," Science, \textbf{305.5685}, 788-792, (2004).

\bibitem{hoffman2007} A.~J.~Hoffman, et al. ``Negative refraction in semiconductor metamaterials," Nature materials, \textbf{6.12}, 946-950, (2007).

\bibitem{fang2009} A.~Fang, T.~Koschny, and C.~M.Soukoulis, ``Optical anisotropic metamaterials: Negative refraction and focusing," Physical Review B, \textbf{79.24}, 245127, (2009).

\bibitem{Litchinitser2008} N.~M.~Litchinitser, et al., ``Negative refractive index metamaterials in optics," Progress in Optics, \textbf{51}, 1-67, (2008).

\bibitem{rosenblatt2015} G.~Rosenblatt, and M.~Orenstein, ``Perfect lensing by a single interface: defying loss and bandwidth limitations of metamaterials," Physical review letters, \textbf{115.19}, 195504, (2015).

\bibitem{zharov2005} A.~A.~Zharov, et al., ``Birefringent left-handed metamaterials and perfect lenses for vectorial fields," New Journal of Physics, \textbf{7.1}, 220, (2005).

\bibitem{zhang2011} X.~Zhang, and S.~R.~Forrest, ``Theory of the perfect lens," Physical Review B, \textbf{84.4}, 045427, (2011).

\bibitem{liu2007}  Z.~S.~ Liu,et al, ``Perfect negative refraction lens using a single homogeneous isotropic metamaterial," Physical review letters, \textbf{99(23)}, 233901, (2007).

\bibitem{schurig2006} D.~Schurig,  J.~B.~ Pendry, and D.~R.~ Smith, ``Metamaterial electromagnetic cloak at microwave frequencies," Science, \textbf{311(5758)}, 1661-1664, (2006).

\bibitem{holloway2005} C.~M.~ Holloway,et al., ``Metamaterials: Microwave applications," Physical Review B, \textbf{72(11)}, 113104, (2005).

\bibitem{cummer2004} S.~A.~ Cummer, et al., ``One-dimensional negative refractive index metamaterials," Nature, \textbf{427(6974)}, 899-903, (2004).

\bibitem{zhu2017}  J.~ Zhu, and A.~ Feng, ``Metamaterial design using machine learning," Materials Today, \textbf{20(11)}, 585-594, (2017).

\bibitem{cai2010} W.~ Cai, et al., ``Metamaterials with negative refractive index at optical wavelengths," Science, \textbf{330(6004)}, 521-525, (2010).

\bibitem{spint1} G.~Armelles, et al., ``Metamaterial platforms for spintronic modulation of mid-infrared response under very weak magnetic field," ACS Photonics, \textbf{5.10}, 3956-3961, (2018).

\bibitem{spint2} A.~Sidorenko, ``Functional Nanostructures and Metamaterials for Superconducting Spintronics," Cham, Switzerland: Springer International Publishing (2018).

\bibitem{spint3} N.~Shitrit, et al., ``Spin-optical metamaterial route to spin-controlled photonics," Science, \textbf{340.6133}, 724-726, (2013).

\bibitem{liu2010} P.~ Liu,et al., ``Bioinspired metamaterials with tunable mechanical properties," Nature materials, \textbf{9(8)}, 694-701, (2010).

\bibitem{tavallaee2010} Amir Ali~ Tavallaee, et al., ``Zero-index terahertz quantum-cascade metamaterial lasers," IEEE Journal of Quantum Electronics, \textbf{46(7)}, 1091-1098, (2010).

\bibitem{ziolkowski2006} Richard W.~Ziolkowski, ``Ultrathin, metamaterial-based laser cavities," JOSA B,\textbf{23(3)}, 451-460, (2006).

\bibitem{boardman2011}Allan D.~ Boardman, et al., ``Active and tunable metamaterials," Laser and Photonics Reviews, \textbf{5(2)}5, 287-307, (2011).

\bibitem{fu2017} Fu~Yangyang, et al., ``Coherent perfect absorber and laser modes in purely imaginary metamaterials," Physical Review A, \textbf{96(4)}, 043838, (2017).

\bibitem{prl-2009} A.~Mostafazadeh, Phys.~ Rev.~ Lett.~\textbf{(102)}, 220402 (2009).

\bibitem{p123} A.~Mostafazadeh, edited by P.~Kielanowski, P.~Bieliavsky, A.~Odzijewicz, M.~Schlichenmaier, and T.~Voronov, ``Geometric Methods in Physics, Trends in Mathematics," Springer, Cham, \textbf{1412.0454}, 145-165,(2015).

\bibitem{naimark} M.~A.~Naimark, Trudy Moscov. Mat. Obsc. \textbf{3}, 181 (1954) in Russian, English translation: Amer. Math. Soc. Transl. (2), \textbf{16}, 103, (1960).

\bibitem{naimark-1} G.~Sh.~Guseinov, Pramana J.~Phys. \textbf{73}, 587, (2009).

\bibitem{CPA} A.~Mostafazadeh, M.~Sarisaman, Phys.~Lett.~A, \textbf{(375)}, 3387, (2011); Proc.~R.~Soc.~Lond.~Ser.~A Math.~Phys.~Eng.~Sci.,~\textbf{(468)}, 3224, (2012); Phys.~ Rev.~ A \textbf{(87)}, 063834, (2013); Phys.~Rev.~ A, \textbf{(88)}, 033810, (2013); Phys.~ Rev.~ A~\textbf{(91)}, 043804, (2015).

\bibitem{Oktay2020}  G.~Oktay, M.~Sarısaman, and M.~Tas, ``Lasing with Topological Weyl Semimetal," Scientific Reports, \textbf{10(1)}, 3127, (2020); doi: 10.1038/s41598-020-59423-3.

\bibitem{mostafazadehmet1} A.~Mostafazadeh, ``Point interactions, metamaterials, and PT-symmetry," Annals of Physics, \textbf{368}, 56-69, (2016).

\bibitem{mostafazadehmet2} F.~Loran, and A.~Mostafazadeh, ``Scattering of Transverse Electric and Transverse Magnetic Waves and Quantum Dynamics Generated by non-Hermitian Hamiltonians," Progress of Theoretical and Experimental Physics, \textbf{2024.12}, 123A01, (2024).

\bibitem{bender} C.~M.~Bender and S.~Boettcher, Phys.~Rev.~Lett. ~\textbf{80}, 5243, (1998); K.~G.~Makris, R.~El-Ganainy, D.~N.~Christodoulides, and Z.~H.~Musslimani, Phys.~Rev.~Lett., ~\textbf{100}, 103904, (2008).

\bibitem{ijgmmp-2010} A.~Mostafazadeh, Int. J. Geom. Meth. Mod. Phys., \textbf{(7)}, 1191, (2010); C.~M.~Bender, D.~C.~Brody, and H.~F.~Jones, Am.~J.~Phys., ~\textbf{(71)}, 1095, (2003).

\bibitem{longhi4} S.~Longhi, Phys. Rev. A, \textbf{(82)}, 032111, (2010).

\bibitem{longhi3} S.~Longhi, J. Phys. A, \textbf{44}, 485302, (2011).

\bibitem{nonhermit1} Yuto Ashida, Zongping Gong, and Masahito Ueda, ``Non-hermitian physics," Advances in Physics, \textbf{69(3)}, 249-435, (2020).

\bibitem{nonhermit2} Kohei Kawabata, et al., ``Symmetry and topology in non-Hermitian physics," Physical Review X, \textbf{9(4)}, 041015, (2019).

\bibitem{nonhermit3} Ramy El-Ganainy, et al, ``Non-Hermitian physics and PT symmetry," Nature Physics, \textbf{14(1)}, 11-19, (2018).

\bibitem{nonhermit4} Nobuyuki Okuma, and Masatoshi Sato, ``Non-Hermitian topological phenomena: A review," Annual Review of Condensed Matter Physics, \textbf{(14)}, 83-107, (2023).

\bibitem{nonhermit5} Hongfei Wang, et al., ``Topological physics of non-Hermitian optics and photonics: a review," Journal of Optics, \textbf{(23.12)}, 123001, (2021).

\bibitem{nonhermit6} Emil J.~Bergholtz, Jan Carl Budich, and Flore K.~ Kunst, ``Exceptional topology of non-Hermitian systems," Reviews of Modern Physics, \textbf{(93.1)}, 015005, (2021).

\bibitem{nonhermit7} Carl M. ~Bender, ``Making sense of non-Hermitian Hamiltonians," Reports on Progress in Physics, \textbf{(70.6)}, 947, (2007).

\bibitem{nonhermit8} Miguel A. ~ Bandres, and Mordechai Segev, ``Non-Hermitian topological systems", Physics, \textbf{(11)}, 96, (2018).

\bibitem{nonhermit9}  Kun Ding, Chen Fang, and Guancong Ma, ``Non-Hermitian topology and exceptional-point geometries," Nature Reviews Physics, \textbf{(4.12)}, 745-760, (2022).

\bibitem{nonhermit10} Nimrod Moiseyev, ``Non-Hermitian quantum mechanics," Cambridge University Press, (2011).

\bibitem{nonhermit11} Ananya Ghatak, and Tanmoy Das, ``New topological invariants in non-Hermitian systems," Journal of Physics: Condensed Matter, \textbf{(31.26)}, 263001, (2019).

\bibitem{nonhermit12} V.~M.~ Martinez Alvarez, et al., ``Topological states of non-Hermitian systems," The European Physical Journal Special Topics, \textbf{(227)}, 1295-1308, (2018).

\bibitem{nonhermit13} Zongping Gong, et al, ``Topological phases of non-Hermitian systems," Physical Review X, \textbf{(8.3)}, 031079, (2018).

\bibitem{lastpaper} A.~Mostafazadeh and M.~Sarisaman, Ann. Phys. (NY),~\textbf{(375)}, 265-287, (2016).

\bibitem{pra-2011a} A.~Mostafazadeh, Phys.~ Rev.~ A, \textbf{(83)}, 045801,(2011).

\bibitem{pra-2012a} A.~Mostafazadeh, Phys.~ Rev.~ A, \textbf{(87)}, 012103,(2012).

\bibitem{hamed2020} Ghaemi-Dizicheh, Hamed, Ali Mostafazadeh, and Mustafa Sarısaman, ``Spectral singularities and tunable slab lasers with 2D material coating," JOSA B, \textbf{(37.7)}, 2128-2138, (2020).

\bibitem{sarisaman20192} M.~Sarısaman, and M.~Tas, ``Broadband coherent perfect absorber with PT-symmetric 2D-materials," Annals of Physics, \textbf{(401)}, 139-148, (2019).

\bibitem{Li2007} J.~ Li, ``Error analysis of mixed finite element methods for wave propagation in double negative metamaterials," Journal of computational and applied mathematics, \textbf{209(1)}, 81-96, (2007)

\bibitem{gainmat1} Lagarkov, Andrey N., Vladimir N. Kisel, and Andrey K. Sarychev. ``Loss and gain in metamaterials," JOSA B, \textbf{27.4},648-659, (2010).

\bibitem{gainmat2} O.~Hess, et al., ``Active nanoplasmonic metamaterials," Nature materials, \textbf{11.7},573-584, (2012).

\bibitem{gainmat3} Wuestner, Sebastian, et al., ``Overcoming losses with gain in a negative refractive index metamaterial," Physical review letters, \textbf{105.12},127401, (2010).

\bibitem{gainmat4} Pacheco-Peña, Victor, and Nader Engheta, ``Temporal metamaterials with gain and loss," arXiv preprint, arXiv:2108.01007 (2021).

\bibitem{Li2008} J.~ Li, Y.~ Chen, and V.~ Elander, ``Mathematical and numerical study of wave propagation in negative-index materials," Computer methods in applied mechanics and engineering, \textbf{197(45-48)}, 3976-3987,(2008). 

\bibitem{Huang2012} Y.~ Huang, J.~ Li, and W.~ Yang, ``Solving metamaterial Maxwell’s equations via a vector wave integro-differential equation," Computers and Mathematics with Applications, \textbf{63(12)}, 1597-1606, (2012).

\bibitem{Rihani2022}  M.~Rihani, ``Maxwell's equations in presence of metamaterials," (Doctoral dissertation, Institut polytechnique de Paris), (2022).

\bibitem{Ziolkowski2003} R.~W.~ Ziolkowski, ``Pulsed and CW Gaussian beam interactions with double negative metamaterial slabs," Opt. Exp., \textbf{(11)}, 662–681, (2003).

\bibitem{absorp1} R.~Avrahamy, et al., ``Chalcogenide-based, all-dielectric, ultrathin metamaterials with perfect, incidence-angle sensitive, mid-infrared absorption: inverse design, analysis, and applications," Nanoscale, \textbf{13.26}, 11455-11469, (2021).

\bibitem{absorp2} J.N.~Monks, et al., ``A wide-angle shift-free metamaterial filter design for anti-laser striking application," Optics Communications, \textbf{429}, 53-59 (2018).

\bibitem{chen2018} Xiaoli Chen,et al., ``Selective metamaterial perfect absorber for infrared and 1.54 laser compatible stealth technology," Optik, \textbf{(172)}, 840-846, (2018).
 
\bibitem{monks2018} James N.~ Monks, et al. ``A wide-angle shift-free metamaterial filter design for anti-laser striking application," Optics Communications, \textbf{(429)}, 53-59, (2018).

\bibitem{ji2021} Xiangbo Ji,et al., ``Perfect metamaterial absorber improved laser-driven flyer," Nanophotonics, \textbf{(10.10)}, 2683-2693, (2021).

\bibitem{hu2016} Xin Hu, et al., ``Metamaterial absorber integrated microfluidic terahertz sensors," Laser and Photonics Reviews, \textbf{(10.6)}, 962-969, (2016).

\bibitem{lin1} Lin, Shirong, et al., ``Controllable flatbands via non-Hermiticity," Applied Physics Letters, \textbf{123.22}, (2023).

\bibitem{lin2} Liang, Yao, et al., ``Hybrid anisotropic plasmonic metasurfaces with multiple resonances of focused light beams," Nano Letters, \textbf{21.20}, 8917-8923, (2021).

\bibitem{CPA2} Y.~ D.~ Chong, L.~ Ge, H.~ Cao, and A.~ D.~ Stone, ``Coherent Perfect Absorbers: Time-Reversed Lasers,” Phys.~ Rev.~ Lett.~ \textbf{(105)}, 053901, (2010); W.~ Wan, Y.~ Chong, L.~ Ge, H.~ Noh, A.~ D.~ Stone, and H.~ Cao, ``Time-Reversed Lasing and Interferometric Control of Absorption,” Science, \textbf{(331)}, 889, (2011); Y.~ D.~ Chong, L.~ Ge, and A.~ D.~ Stone, ``PT-Symmetry Breaking and Laser-Absorber Modes in Optical Scattering Systems,” Phys.~ Rev.~ Lett.~, \textbf{(106)}, 093902, (2011); S.~ Longhi, ``PT-symmetric laser absorber,” Phys.~ Rev.~ A, \textbf{(82)}, 031801, (2010); ``Coherent perfect absorption in a homogeneously broadened two-level medium,” Phys.~ Rev.~ A, \textbf{(83)}, 055804, (2011); and ``Time-Reversed Optical Parametric Oscillation,” Phys.~ Rev.~ Lett.~, \textbf{(107)}, 033901, (2011); L.~ Ge, Y.~D.~Chong, S.~ Rotter, H.~ E.~ Türeci, and A.~D.~ Stone, "Unconventional modes in lasers with spatially varying gain and loss,” Phys.~  Rev.~  A, \textbf{(84)}, 023820, (2011).

\bibitem{ct1} Y.~Liang, P.~T.~Din and Y.~Kivshar, ``From local to nonlocal high-Q plasmonic metasurfaces," Physical Review Letters, \textbf{133.5}, 053801, (2024).

\bibitem{plasma} I.~Levchenko, S.~Xu, O.~Cherkun, O.~Baranov, and K.~Bazaka, ``Plasma meets metamaterials: Three ways to advance space micropropulsion systems," Advances in Physics: X, \textbf{6(1)}, (2020).























\end{thebibliography}
\end{document}